\documentclass[preprint2]{aastex}

\usepackage{graphicx}
\DeclareGraphicsExtensions{.ps}
\shorttitle{\indent \def Reconnection in earliest phase of flux emergence} \shortauthors{Tian et al.}

\begin{document}

\title{Magnetic reconnection at the earliest stage of solar flux emergence}

\author{Hui Tian\altaffilmark{1}, Xiaoshuai Zhu\altaffilmark{2,3}, Hardi Peter\altaffilmark{3}, Jie Zhao\altaffilmark{4}, Tanmoy Samanta\altaffilmark{1}, Yajie Chen\altaffilmark{1}}
\altaffiltext{1}{School of Earth and Space Sciences, Peking University, Beijing 100871, China; huitian@pku.edu.cn}
\altaffiltext{2}{Key Laboratory of Solar Activity, National Astronomical Observatories, Chinese Academy of Sciences, Beijing 100012, China.}
\altaffiltext{3}{Max Planck Institute for Solar System Research, Justus-von-Liebig-Weg 3, 37077 G\"ottingen, Germany.}
\altaffiltext{4}{Key Laboratory of Dark Matter and Space Astronomy, Purple Mountain Observatory, CAS, Nanjing 210008, China.}

\begin{abstract}

On 2016 September 20, the Interface Region Imaging Spectrograph observed an active region during its earliest emerging phase for almost 7 hours. The Helioseismic and Magnetic Imager on board the Solar Dynamics Observatory observed continuous emergence of small-scale magnetic bipoles with a rate of $\sim$10$^{16}$ Mx~s$^{-1}$. The emergence of magnetic fluxes and interactions between different polarities lead to frequent occurrence of ultraviolet (UV) bursts, which exhibit as intense transient brightenings in the 1400 \AA{} images. In the meantime, discrete small patches with the same magnetic polarity tend to move together and merge, leading to enhancement of the magnetic fields and thus formation of pores (small sunspots) at some locations. The spectra of these UV bursts are characterized by the superposition of several chromospheric absorption lines on the greatly broadened profiles of some emission lines formed at typical transition region temperatures, suggesting heating of the local materials to a few tens of thousands of kelvin in the lower atmosphere by magnetic reconnection. Some bursts reveal blue and red shifts of $\sim$100~km~s$^{-1}$ at neighboring pixels, indicating the spatially resolved bidirectional reconnection outflows. Many such bursts appear to be associated with the cancellation of magnetic fluxes with a rate of the order of $\sim$10$^{15}$ Mx~s$^{-1}$. We also investigate the three-dimensional magnetic field topology through a magneto-hydrostatic model and find that a small fraction of the bursts are associated with bald patches (magnetic dips). Finally, we find that almost all bursts are located in regions of large squashing factor at the height of $\sim$1 Mm, reinforcing our conclusion that these bursts are produced through reconnection in the lower atmosphere. 

\end{abstract}
\keywords{Sun: sunspots---Sun: chromosphere---Sun: transition region---Sun: UV radiation---magnetic reconnection}

\section{Introduction}

As one of the most important mechanisms of energy release, magnetic reconnection plays an important role in the energization of the space and astrophysical plasmas \citep[e.g.,][]{Priest2000,Deng2004,Gosling2005,Phan2006,Zhang2012}. Magnetic reconnection is also believed to be the key process that drives both large-scale solar eruptions and small-scale jets \citep[e.g.,][]{Sun2015,Chen2015,Sterling2015,Chen2016,Xue2016,Li2016,Wyper2017,Ni2017}. In the past decade, due to high-resolution observations by advanced telescopes such as the Solar Optical Telescope \citep[SOT,][]{Tsuneta2008} on board the Hinode spacecraft, the Interface Region Imaging Spectrograph \citep[IRIS,][]{DePontieu2014} and the New Vacuum Solar Telescope \citep[NVST,][]{Liu2014}, small-scale reconnection in the partially ionized lower solar atmosphere has received a lot of attention \citep[e.g.,][]{Shibata2007,Katsukawa2007,Tian2014a,Peter2014,Yang2015}.

One type of such reconnection events is the so-called hot explosion in the cool solar atmosphere \citep{Peter2014}. These events are revealed as compact intense brightenings in the transition region images obtained with the 1400 \AA{} and 1330 \AA{} filters of IRIS. Such transient brightenings are also called IRIS bombs or ultraviolet (UV) bursts by different authors. We will use the term UV burst to describe this phenomenon in the rest of this paper. The IRIS spectra of these events are characterized by the superposition of several chromospheric absorption lines (mainly from the ions of Ni~{\sc{ii}} and Fe~{\sc{ii}}) on the greatly broadened profiles of several emission lines from the Si~{\sc{iv}} and C~{\sc{ii}} ions, indicating heating of the local materials to a temperature of a few tens of thousands of kelvin in the lower chromophere or even the photosphere \citep{Peter2014,Rutten2016}. The greatly enhanced wings of the Si~{\sc{iv}}~1393.755 \AA{}~and 1402.770 \AA{}~lines are often believed to be caused by the superposition of bi-directional reconnection outflows in the line of sight \citep[e.g.,][]{Innes1997,Peter2014}, although \cite{Judge2015} proposed that they are more compatible with Alfv\'enic turbulence. Enhanced line wings have also been found in some spatially resolved bi-directional jets \citep[e.g.,][]{Vissers2015}, possibly related to the flow inhomogeneity or turbulence in the outflow regions. Signatures of photospheric flux cancellation associated with some UV bursts also lend support to the cause of these events by magnetic reconnection \citep{Peter2014,Zhao2017}.

The significant heating of UV bursts has received great attention in the past three years. Combined IRIS and ground-based observations by \cite{Vissers2015}, \cite{Kim2015} and \cite{Tian2016} show that some of the UV bursts are closely related to the well-known Ellerman bombs (EBs), which are characterized by intense transient brightening of the extended wings of the H$_{\alpha}$ line \citep[e.g.,][]{Ellerman1917,Ding1998,Georgoulis2002,Watanabe2008,Watanabe2011,Yang2013,Vissers2013,Vissers2015,Nelson2013,Nelson2015}. EBs are usually found in emerging  active regions (ARs). However, recent observations by \cite{Rouppe2016} and \cite{Nelson2017} found EB-like brightenings also in the quiet Sun. EBs are believed to be generated by magnetic reconnection in the photosphere or around the temperature minimum region (TMR) \citep[e.g.,][]{Watanabe2011,Vissers2013,Nelson2015,Rezaei2015,Danilovic2017,Chen2017}. Thus, the association of some UV bursts with EBs suggests that reconnection around the TMR may heat the cool materials to a temperature of $\sim$80,000 K, the formation temperature of the Si~{\sc{iv}}~lines under ionization equilibrium. A recent investigation of EB emission signals in the He~{\sc{i}} D${_3}$ and 10830 \AA{} lines also indicates a similar temperature enhancement \citep{Libbrecht2017}. Such a scenario appears to challenge existing models of EBs, in which the TMR could not be heated to a temperature higher than 10,000 K \citep[e.g.,][]{Fang2006,Fang2017,BelloGonzalez2013,Berlicki2014,Hong2014,Hong2017a,Hong2017b,Li2015,Reid2017}. It is worth noting that \cite{Rutten2016} put the formation of the Si~{\sc{iv}}~lines at a temperature of 10,000--20,000 K by assuming local thermodynamic equilibrium for line extinctions during the onsets of EBs. This temperature is still higher than the model predictions. Recently three-dimensional radiative magnetohydrodynamic simulations by \cite{Hansteen2017} have successfully produced EBs and UV bursts. However, in their simulations these two types of events occur at different heights and different times, thus difficult to explain the well observed connection between UV bursts and EBs \citep{Vissers2015,Kim2015,Tian2016}. 

IRIS observations show that UV bursts tend to occur in emerging active regions. Indeed, numerical simulations of magnetic flux emergence and AR formation have revealed sporadic small-scale reconnection events \citep[e.g.,][]{Isobe2007,Archontis2009,Pariat2009,Cheung2010}. The local heating resulting from these reconnection events was proposed to explain EBs. Given the obvious connection between a significant fraction of UV bursts and EBs \citep{Tian2016}, the reconnection scenario in these models may be applicable to at least some of the UV bursts as well. Investigations of these local heating events at different stages of flux emergence can improve our understanding of not only the interaction between different parts of the emerging flux system, but also the evolution of the magnetic field topology during the formation of active regions. However, most existing IRIS observations of UV bursts were performed during relatively late stages of flux emergence (more than 6 hours after the initial appearance of sunspots or pores). Most recently, \cite{Toriumi2017} studied various local heating events appearing ~2 hours after the start of flux emergence in a developing AR and obtained crucial information about the magnetic field topologies of these events. To investigate the dependence of these heating events on the evolution of ARs, more IRIS observations of the earliest-stage flux emergence need to be performed.

On 2016 September 20, IRIS observed an AR during its earliest emerging phase for almost 7 hours, starting from the first sign of flux emergence. Here we present analysis results of the temporal evolution of the intense local heating events in this observation, which provides unique information on the reconnection processes at the earliest stage of flux emergence and AR formation.

\section{Observations}

\begin{figure*}
\centering {\includegraphics[width=\textwidth]{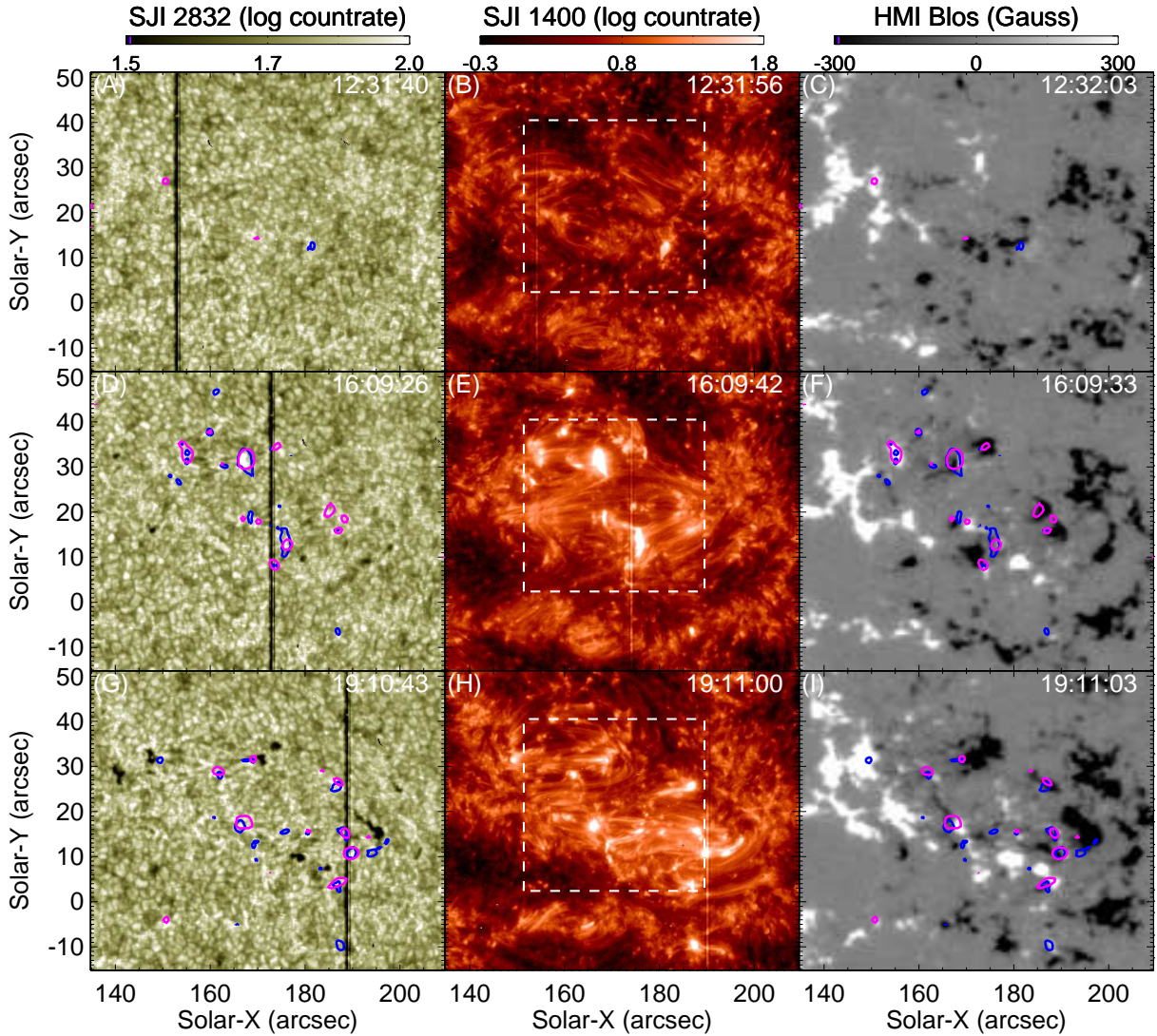}} \caption{ IRIS/SJI 2832 \AA{}~images, 1400 \AA{} images and HMI line-of-sight magnetograms taken around 12:32 UT, 16:09 UT and 19:11 UT. Blue/purple contours marking the compact brightenings in the 1400 \AA{}/1700 \AA{} images are overplotted in the 2832 \AA{}~images and the magnetograms. A movie (m1.mov) showing the complete sequences of these images are available online. The rectangles shown in the middle and right panels outline the regions used for the calculation of the intensities and magnetic fluxes shown in Figure~\ref{fig.2}. } \label{fig.1}
\end{figure*}

IRIS performed 23 large sparse 64-step rasters (120$^{\prime\prime}$ along the slit, 64 raster steps with a step size of $\sim$1$^{\prime\prime}$) in a region to the west of NOAA active region (AR) 12593 from 12:24 UT to 19:11 UT. Each raster lasted for 1061 seconds. The target was close to the disk center, with a pointing coordinate of (158$^{\prime\prime}$, 2$^{\prime\prime}$). The cadence of the spectral observation was $\sim$16.6 seconds (time difference between two consecutive exposures), with an exposure time of 15 seconds. Slit-jaw images (SJI) in the 2832 \AA{}~(mainly Mg~{\sc{ii}} wing emission from the photosphere), 2796 \AA{}~(mainly the Mg~{\sc{ii}} k line), 1330 \AA{}~(mainly ultraviolet continuum and two C~{\sc{ii}} lines) and 1400 \AA{}~(mainly ultraviolet continuum and two Si~{\sc{iv}} lines) filters were taken alternately with a cadence of $\sim$66 seconds for each filter. The spatial pixel size of the images is $\sim$0.166$^{\prime\prime}$. The far and near ultraviolet wavelength bands have a spectral dispersion of $\sim$0.013  \AA{} and $\sim$0.025  \AA{} per pixel, respectively. Dark current subtraction, flat fielding, geometrical and orbital variation corrections have been applied in the level 2 data used here. The fiducial lines are used to achieve an alignment between images taken in different spectral windows and SJI filters.

We have also performed absolute wavelength calibration for the IRIS spectra. The chromospheric Fe~{\sc{ii}}~1392.817 \AA{} line is used for the wavelength calibration of the Si~{\sc{iv}}~1393.755 \AA{}~spectral window. We assume a zero Doppler shift for the profile of the Fe~{\sc{ii}}~1392.817 \AA{} line averaged over all spatial pixels and all exposures in the first raster. This cold line is known to have a negligible velocity on average \citep{Tian2014a,Peter2014}. For the wavelength calibration of the Si~{\sc{iv}}~1402.770 \AA{}~window, we simply assume the same Doppler shift of the two Si~{\sc{iv}}~lines in the averaged spectra of the first raster. The wavelength calibration for the C~{\sc{ii}}~window has been achieved by forcing the Ni~{\sc{ii}}~1335.203 \AA{} and 1393.330 \AA{} lines to have the same Doppler shift in the same average spectra. Note that here we do not use the O~{\sc{i}}~1355.60 \AA{} line as originally suggested in the IRIS technical note 20, since this line is too far from the strong Si~{\sc{iv}}~and C~{\sc{ii}}~lines. The wavelength calibration in the Mg~{\sc{ii}} window is performed by assuming a zero shift for several strong neutral absorption lines present in the reference spectra. For identification of these lines we refer to \cite{Tian2017}.

We have also analyzed images taken in the 1700 \AA{}~passband of the Atmospheric Imaging Assembly \citep[AIA,][]{Lemen2012} instrument on board the Solar Dynamics Observatory \citep[SDO,][]{Pesnell2012}. The 1700 \AA{}~passband mainly samples the ultraviolet continuum emission formed around the TMR. The cadence and pixel size of the 1700 \AA{}~images are 24 s and $\sim$0.6$^{\prime\prime}$, respectively. After the AIA 1700 \AA{}~and IRIS 1400 \AA{}~images are both internally aligned through the technique of cross correlation, we coalign images taken in these two passbands by checking locations of some commonly observed transient brightenings by eye. To investigate the magnetic field structures associated with UV bursts, we have also analyzed the data taken by the Helioseismic and Magnetic Imager \citep[HMI,][]{Scherrer2012} on board SDO. The cadence of the line-of-sight magnetograms taken by HMI is 45 seconds. The pixel size is $\sim$0.5$^{\prime\prime}$. Once the HMI magnetograms are internally aligned through cross correlation, the coalignment between IRIS images and HMI magnetograms is achieved by matching the bright network lanes in 1400 \AA{}~images and the flux concentrations in HMI magnetograms at the beginning of the IRIS observation.

\section{History of flux emergence}

\begin{figure*}
\centering {\includegraphics[width=0.8\textwidth]{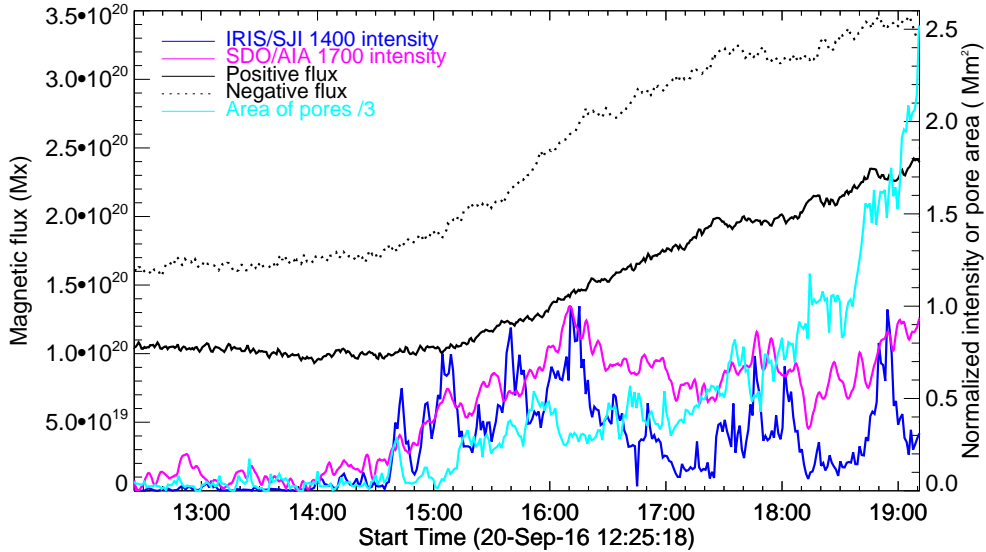}} \caption{ Temporal evolution of the AIA 1700 \AA{} intensity, SJI 1400 \AA{} intensity, total area of pores, positive and negative magnetic fluxes. Each of the two intensity (countrate) light curves has been normalized to the maximum intensity in the time series). The area of pores has been divided by 3 for the purpose of illustration. } \label{fig.2}
\end{figure*}

Figure~\ref{fig.1} presents the HMI line-of-sight magnetograms, IRIS/SJI 2832 \AA{}~and 1400 \AA{} images taken around 12:32 UT, 16:09 UT and 19:11 UT. Only part of the full field of view of the IRIS observation is shown here. The flux emergence process is clearly revealed in Figure~\ref{fig.1} and the associated online movie. At the beginning of the IRIS observation, the target appears to be a typical quiet-Sun region, showing obvious network structures in the magnetograms~and 1400 \AA{}~ images. The 2832 \AA{}~images reveal essentially only granules. However, a close inspection of the magnetograms suggests the first sign of flux emergence around the location of (183$^{\prime\prime}$, 14$^{\prime\prime}$), where compact brightenings are observed in the 1400 \AA{} images. The small loop system connecting the emerging bipoles can be clearly identified from the 1400 \AA{} images about 1 hour later. As time evolves, more small-scale magnetic bipoles emerge. Meanwhile, more transient brightenings appear in the 1400 \AA{} images and they seem to be related to the interactions between fluxes with different polarities. Presumably, these intense brightenings are UV bursts. The sequence of magnetograms also reveals a clear trend that discrete small patches with the same polarity move together and merge, leading to enhancement of the magnetic field strength and thus formation of pores (the darkest features in the 2832 \AA{}~images) at some locations. The emerging loop complex can be identified from the 1400 \AA{}~images. We have also examined the AIA 304 \AA{} and 171 \AA{} images, and some UV bursts appear to show up in the 304 \AA{} images. A very small fraction of the bursts are also visible in the 171 \AA{} images, which might be caused by the transition region contribution to the passband. Some emerging loops could also be identified from the AIA 304 \AA{} and 171 \AA{} images.

The history of flux emergence is presented in Figure~\ref{fig.2}, where the temporal evolution of the total positive and negative magnetic fluxes in the central emerging flux region (Solar-X=[151.4$^{\prime\prime}$ -- 189.5$^{\prime\prime}$], Solar-Y=[2.4$^{\prime\prime}$ -- 40.5$^{\prime\prime}$]) is plotted. The same region is also used to calculate the intensity (in countrate) integrated over all pixels where the countrate is larger than 38 (or the common logarithm of countrate is larger than 1.58, within the blue contours) in each 1400 \AA{} image. When all time steps are considered, the total number of such pixels is 1\% of the total number of pixels in this central emerging flux region. Previous studies showed that some UV bursts are clearly visible in the 1700 \AA{} images \citep{Vissers2015,Tian2016}. This is also the case in our observation (see Figure~\ref{fig.1} and the associated online animation). Thus, for each 1700 \AA{} image we also calculate the intensity (in countrate) integrated over all pixels where the countrate is larger than 3939 (within the purple contours) in the same rectangular region. Again the total number of these pixels is 1\% of the total pixel number in this rectangular region when all time steps are counted. The total areas of pores at different times are also calculated from the 2832 \AA{}~images by using an intensity threshold of 35.5 (countrate) or 1.55 (common logarithm of countrate). Since the western part of the field of view shown in Figures~\ref{fig.1} was not observed at some occasions due to the slit scanning, we only take the part of the 2832 \AA{}~images in the range of Solar-X=[134.7$^{\prime\prime}$ -- 186.3$^{\prime\prime}$] for the calculation of the pore areas.  Figure~\ref{fig.2} shows a dramatic increase of the 1700 \AA{} and 1400 \AA{} intensities around 14:40 UT, indicating the occurrence of many UV bursts. Such intense local heating is clearly related to the substantial emergence of magnetic fluxes. Slightly after that, we see a significant increase of the pore area.

After 15:00 UT, the magnetic fluxes still keep emerging and the pore area also keeps increasing. However, the 1700 \AA{} intensity appears to stay at a relatively stable level, suggesting that the heating by UV bursts is more or less constant over a few hours. As pointed out by \cite{Rutten2016}, the AIA 1700 \AA{} channel is dominated by the Balmer continuum at high temperatures. Considering the magnetic origin of the energy, the heating rate should be related to the injection rate of magnetic fluxes. We thus expect a constant rate of flux emergence, which agrees with our observation as the magnetic fluxes increase almost linearly from 15:00 UT to 19:11 UT. During this period the flux of either polarity has increased by $\sim$1.5$\times$10$^{20}$ Mx, as shown in Figure~\ref{fig.2}. Thus, the flux emergence rate is estimated to be $\sim$10$^{16}$ Mx~s$^{-1}$ for either polarity. The 1400 \AA{} intensity is more variable compared to that of 1700 \AA{}, which is expected as the 1400 \AA{} filter samples mainly the Si~{\sc{iv}} lines. Intensities of such transition region lines are known to be highly variable and very sensitive to small pertubation. Nevertheless, we can still see a general correlation between the 1700 \AA{} and 1400 \AA{} intensities, suggesting that both of them are good indicators of local heating.
%The nearly constant 1700 \AA{} intensity may be understood as a nearly constant heating rate.

\section{Spectra of UV bursts}

\begin{figure*}
\centering {\includegraphics[width=\textwidth]{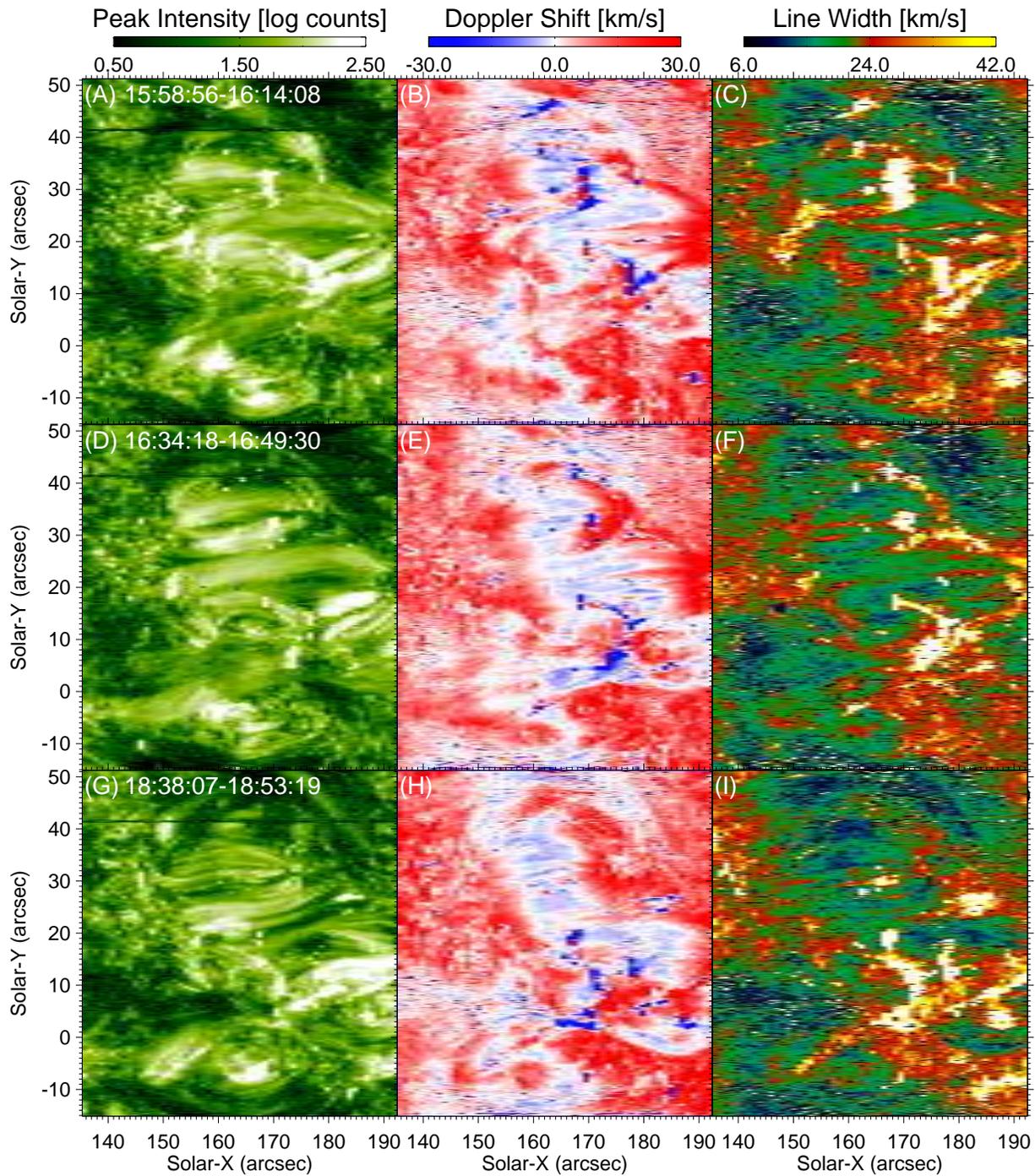}} \caption{ Images of the peak intensity, Doppler shift and line width obtained from a single Gaussian fit to the Si~{\sc{iv}}~1393.755 \AA{}~line profiles acquired during three raster scans. The time ranges during which these spectra were taken are shown in the intensity images. } \label{fig.3}
\end{figure*}

We first apply a single Gaussian fit to the line profiles of Si~{\sc{iv}}~1393.755 \AA{}~acquired during all the 23 rasters. At locations of UV bursts the line profiles are often complex or even multi-peaked. A Gaussian fitting to such line profiles could be meaningless. However, at other locations the line profiles usually can be well fitted with a Gaussian function. The emerging process of the AR is clearly revealed in the image sequences of Si~{\sc{iv}} line parameters. As exemplified in Figure~\ref{fig.3}, these emerging loops can be easily identified from the Si~{\sc{iv}}~intensity images. Such cool loops belong the category of cool transition region loops reported by \cite{Huang2015}. Figure~\ref{fig.3} also reveals significant red shifts on the order of 30~km~s$^{-1}$ at the loop footpoints, indicating the presence of downflows possibly resulting from cooling of the hotter plasma in the loops. While the loop tops are generally blueshifted by a few~km~s$^{-1}$, indicating the slow rising motion of the emerging loops. Such a velocity pattern appears to be consistent with models of flux emergence \citep[e.g.,][]{Chen2014}.

\begin{figure*}
\centering {\includegraphics[width=\textwidth]{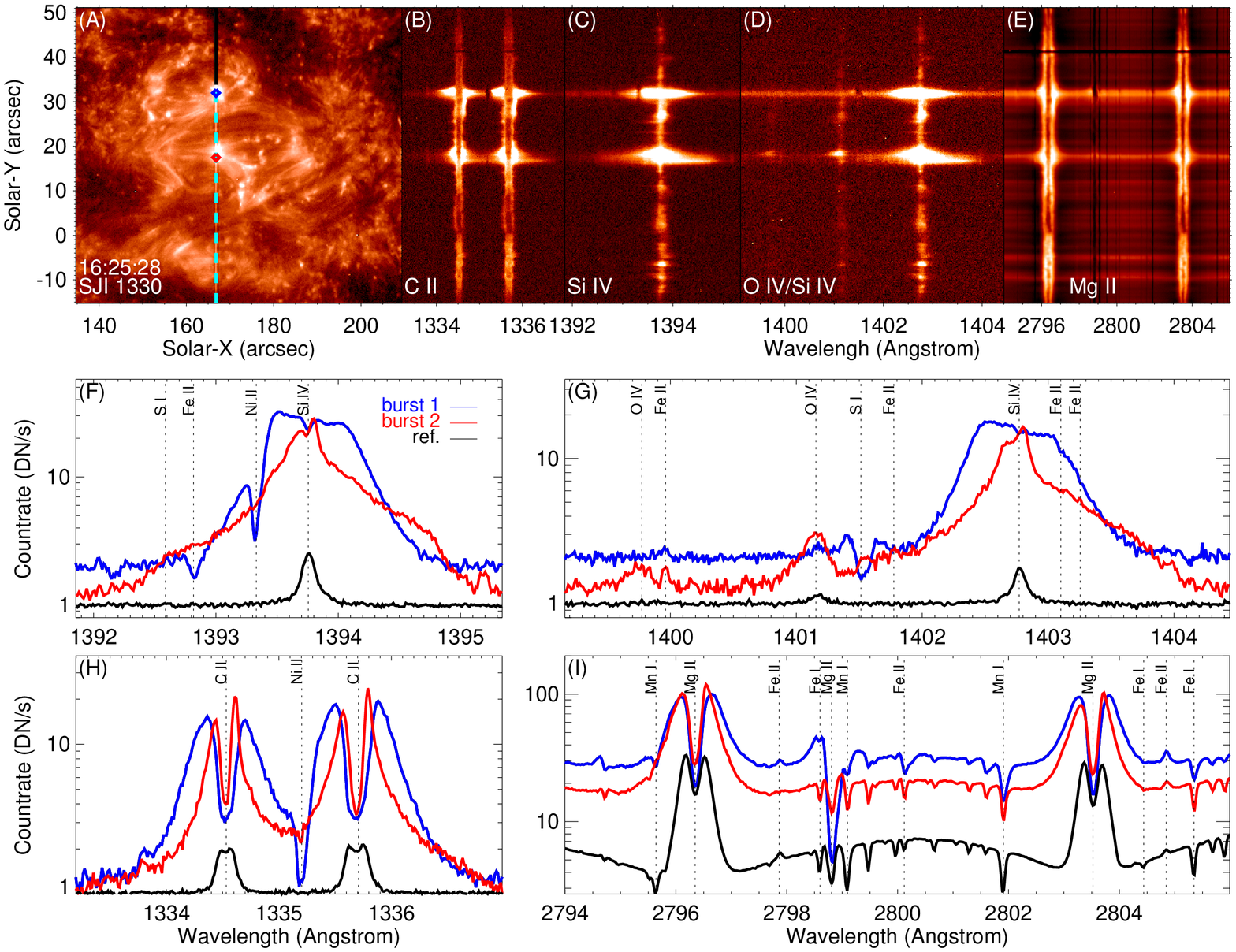}} \caption{ (A)-(E) IRIS 1330 \AA{} image and spectra in four spectral windows taken at 16:25:28 UT. The cyan dashed line in (A) marks the slit position. Two UV bursts are indicated by the blue (burst 1) and red (burst 2) diamonds in (A). (F)-(I) IRIS spectra of bursts 1 and 2 in four spectral windows. The reference spectra (black) are obtained by averaging the line profiles within the section marked by the black line in (A). Rest wavelengths of several spectral lines are indicated by the vertical dashed lines.} \label{fig.4}
\end{figure*}

Discrete patches of significant Doppler shift or adjacent blue shift/red shift can also be identified from the Dopplergrams shown in Figure~\ref{fig.3}. These patches are mostly locations of UV bursts and they appear to be related to the largest line widths. In many of these cases the Si~{\sc{iv}} line profiles are very complex, and thus talking about the Doppler shift and nonthermal line width might not be meaningful. Instead, we decide to discuss characteristics of individual line profiles. Figure~\ref{fig.4} shows the spectra of two UV bursts (bursts 1 and 2) observed at 16:25:28 UT. The spectra of burst 1 are clearly characterized by superposition of several absorption lines on the greatly enhanced and broadened profiles of the Si~{\sc{iv}}, C~{\sc{ii}} and Mg~{\sc{ii}} lines. The absorption lines are mainly from the ions of Fe~{\sc{ii}}~and Ni~{\sc{ii}}, usually revealing a blue shift of a few km~s$^{-1}$. Such spectra suggest the presence of hotter materials (up to the Si~{\sc{iv}} formation temperature $\sim$8$\times$10$^{4}$ K) below the slowly expanding upper chromosphere \citep{Peter2014}. 

As suggested in many previous investigations, the greatly enhanced emission at both wings of some transition region line profiles may result from spatially unresolved bidirectional outflows from a small reconnection region \citep[e.g.,][]{Innes1997,Peter2014}. In quiet-Sun regions line profiles with this characteristics are frequently reported and they are often called transition region explosive events \citep[e.g.,][]{Dere1989,Innes1997,Chae1998,Madjarska2004,Ning2004,Teriaca2004,Huang2014,Gupta2015}. For the UV bursts identified in ARs, IRIS observations often reveal much stronger wing enhancement and line broadening \citep[e.g.,][]{Peter2014,Yan2015,Kim2015,Tian2016}, which may be related to faster reconnection outflows and higher temperatures as a result of the larger amount of released magnetic energy during reconnection. It is also possible that these broad line profiles are caused by the superposition of spatially unresolved plasmoids with different velocities in the reconnection region \citep[e.g.,][]{Innes2015,Rouppe2017}. 

The two UV bursts shown in Figure~\ref{fig.4} appear to have different types of line profiles. The spectra of burst 1 reveals the following distinct characteristics: (1) the chromospheric Ni~{\sc{ii}}~1393.330 \AA{} and 1335.203 \AA{}~lines show a very deep absorption; (2) the forbidden lines O~{\sc{iv}}~1401.156 \AA{} and 1399.774 \AA{} are almost absent; (3) the S~{\sc{i}}~1401.514 \AA{} line reveals enhanced wings bridged by a central depression; (4) the Mn~{\sc{i}}~2795.640 \AA{} absorption line is superimposed on the enhanced wing of the Mg~{\sc{ii}}~k line. These features are typical for the line profiles of EB-related UV bursts \citep{Tian2016}, indicating that burst 1 is likely formed in a very deep and dense layers of the solar atmosphere, i.e., the photosphere. On the contrary, in burst 2 the Ni~{\sc{ii}}~and Mn~{\sc{i}}~absorption lines are very shallow or not visible. Also the O~{\sc{iv}}~lines clearly show up in burst 2. Such line profiles are similar to those of the UV bursts that are not connected to EBs \citep{Tian2016}, indicating that burst 2 is likely formed higher up and probably in the chromopshere.

\begin{figure}
\centering {\includegraphics[width=0.45\textwidth]{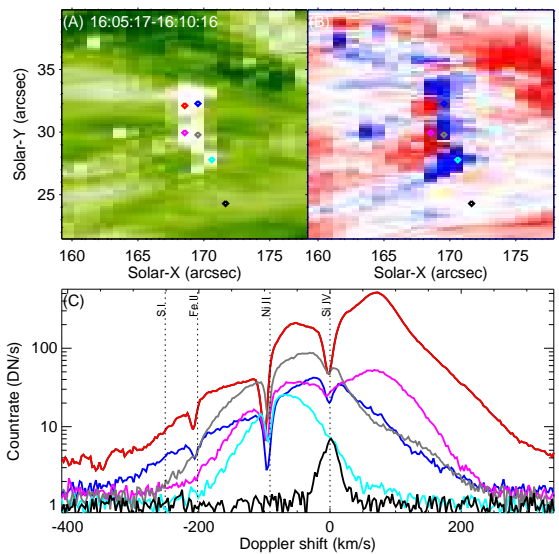}} \caption{ (A)-(B) Images of the peak intensity and Doppler shift of the Si~{\sc{iv}}~1393.755 \AA{}~line obtained from 16:05:17 UT to 16:10:16 UT. (C) The Si~{\sc{iv}}~1393.755 \AA{}~line profiles at the locations marked by the six diamonds with different colors, in and around burst 1. Note that the spectral windows have the same extent in Figure~\ref{fig.4} and Figure~\ref{fig.5}.  } \label{fig.5}
\end{figure}

Burst 1 appears to recur at the same location many times during a period of a few hours. The Si~{\sc{iv}}~Dopplergrams reveal significant blue shifts and red shifts adjacent to each other at this location, suggesting that the two reconnection outflows may have been spatially resolved. Figure~\ref{fig.5} shows the Si~{\sc{iv}}~intensity and Dopplershift images obtained during one raster scan. The line profiles at the redshift and blueshift patches do peak at the red and blue sides of the rest position, respectively. However, most of these line profiles are not entirely shifted. Instead, they generally show asymmetric enhancement at the two wings, leading to a net blue shift or red shift when a single Gaussian fit is applied. We also notice that most Si~{\sc{iv}}~line profiles within the burst show an obvious dip at the rest wavelength of the line. This feature appears to be caused by self-absorption, which is supported by the observed intensity ratio of the Si~{\sc{iv}}~1393.755 \AA{}~and 1402.77 \AA{}~lines. Under optically thin conditions the expected ratio of these two lines is equal to the ratio of the oscillator strength, which is 2 for these Li-like resonance lines \citep[see a discussion in][]{Peter2014}. Within the burst 1 we observe a ratio of $\sim$1.8, which is lower than the observed ratio of $\sim$2 in the surrounding quiet regions and indicates that the Si~{\sc{iv}}~lines become optically thick during the occurrence of the burst. This central dip is more obvious in the stronger Si~{\sc{iv}}~1393.755 \AA{} line, which is also expected if the opacity effect comes into play \citep{Yan2015}. \cite{Yan2015} proposed that such a central dip may be caused by absorption of the overlying transition region loops, which may also be the case in our observations.

\begin{figure*}
\centering {\includegraphics[width=\textwidth]{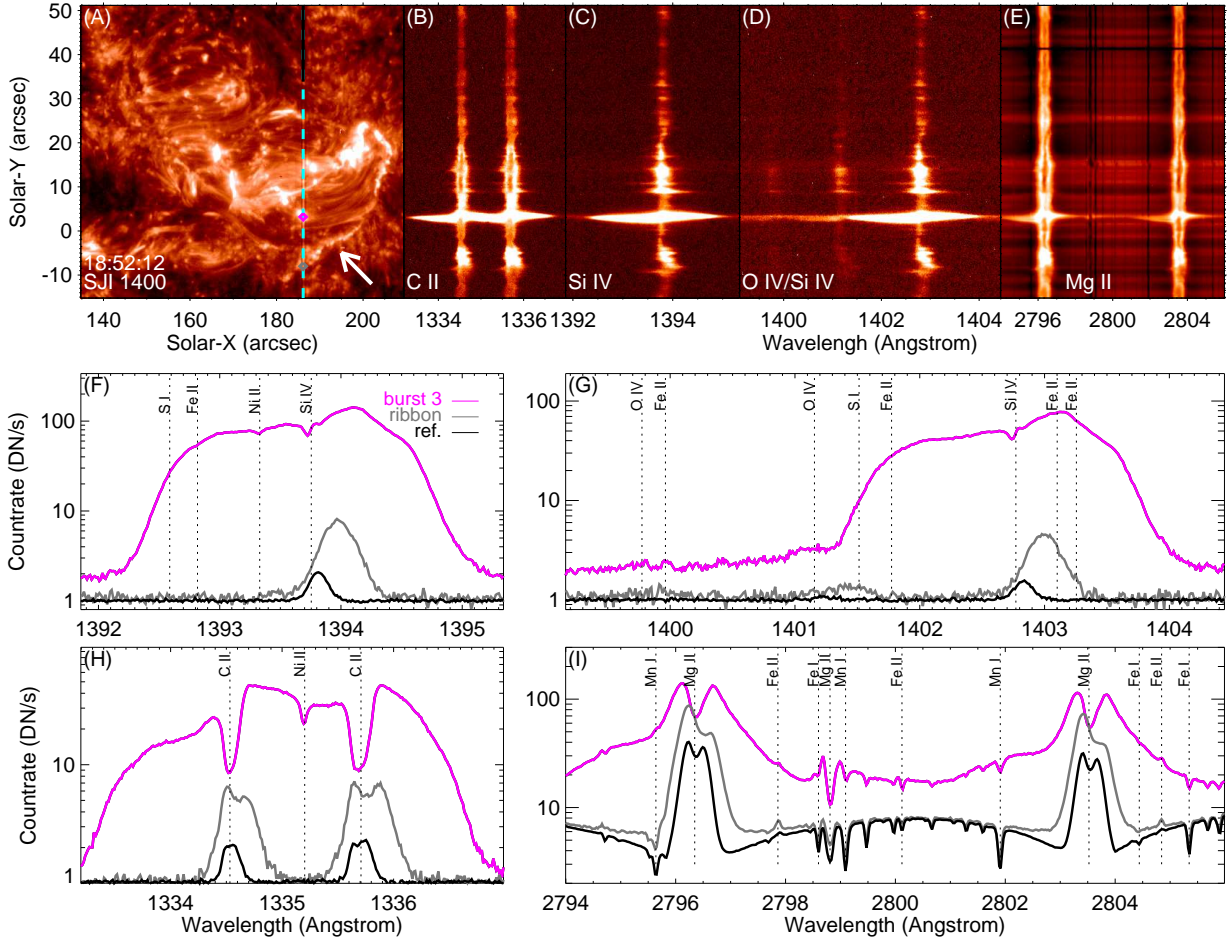}} \caption{ (A)-(E) IRIS 1400 \AA{} image and spectra in four spectral windows taken at 18:52:12 UT. The cyan dashed line in (A) marks the slit position. The purple and grey diamonds indicate a UV burst (burst 3) and the ribbon-like structure (also pointed by the arrow) respectively in (A). (F)-(I) IRIS spectra of burst 3 and the ribbon in four spectral windows. The reference spectra (black) are obtained by averaging the line profiles within the section marked by the black line in (A). Rest wavelengths of several spectral lines are indicated by the vertical dashed lines. } \label{fig.6}
\end{figure*}

\begin{figure}
\centering {\includegraphics[width=0.45\textwidth]{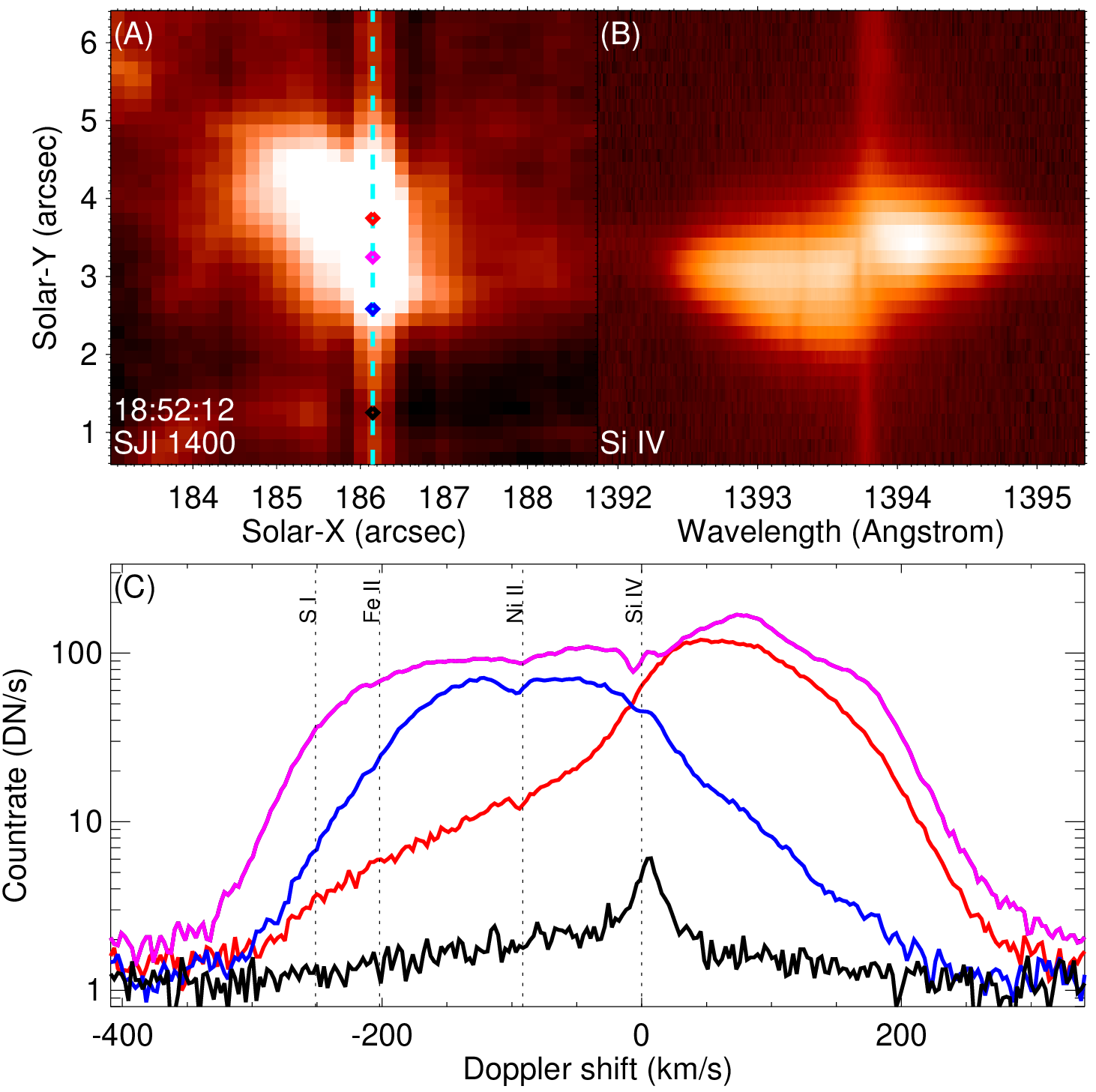}} \caption{ (A)-(B) IRIS 1400 \AA{} SJI image and spectral image of the Si~{\sc{iv}}~1393.755 \AA{}~line taken at 18:52:12 UT. (C) The Si~{\sc{iv}}~1393.755 \AA{}~line profiles at the locations marked by the four diamonds with different colors, in and around burst 3. Note that the spectral windows have the same extent in Figure~\ref{fig.6} and Figure~\ref{fig.7}. } \label{fig.7}
\end{figure}

The IRIS spectra of another UV burst observed at 18:52:12 UT (burst 3) are presented in Figures~\ref{fig.6}. The Si~{\sc{iv}}, C~{\sc{ii}} and Mg~{\sc{ii}} line profiles are significantly broadened, and their significantly enhanced wings extend to at least $\sim$300~km~s$^{-1}$ from the line cores. The peak intensities of the two Si~{\sc{iv}} lines increase by two orders of magnitude. While the integrated line intensities are enhanced by almost three orders of magnitude. Such a significant enhancement has been rarely reported, possibly in only a few previous observations \citep[e.g.,][]{Yan2015,Vissers2015}. The greatly enhanced Mg~{\sc{ii}} wings and the NUV continuum suggest that this burst may be connected to an EB occurring in the photosphere \citep{Tian2016,Grubecka2016}. A closer inspection of the spectra suggests that the signs of Doppler shift are different at the southern and northern parts of the UV burst. Figure~\ref{fig.7}~shows that the line profiles at the southern and northern parts of the UV burst are entirely blueshifted and redshfited, respectively. This is a strong evidence that the bidirectional reconnection outflows are spatially resolved. A similar pattern of Doppler shift has also been recently found for a UV burst analyzed by \cite{Chitta2017}. Greatly broadened line profiles resulting from a mixture of the bidirectional flows can be found in the middle part of burst 3. Such behavior suggests that the reconnection outflows propagate in a plane defined by the slit and the solar surface normal. The paths of the bidirectional outflows are inclined with respect to the line of sight. Figure~\ref{fig.7}~indicates that the bidirectional flows have an average velocity of $\sim$100~km~s$^{-1}$, suggesting that the Alv\'en speed in the reconnection region is of the same order.

It is generally believed that UV bursts are markers of reconnection related local heating. Observations of UV bursts, such as the one discussed here, provide important constraint to numerical simulations of magnetic reconnection in the partially ionized lower solar atmosphere \citep[e.g.,][]{Murphy2015,Ni2015,Ni2018}. Our observations confirm that some UV bursts are likely related to magnetic reconnection around the TMR. Such a scenario has not been reproduced in most theoretical investigations, in which the TMR is usually heated to a temperature less than 10,000 K \citep[e.g.,][]{Fang2006,Fang2017,BelloGonzalez2013,Berlicki2014,Hong2014,Hong2017a,Hong2017b,Li2015,Reid2017}. To solve the discrepancy between theories and observations, \cite{Fang2017} recently proposed two new possibilities. The first possibility is that the heating occurs in the upper atmosphere, through waves or shocks generated by reconnection jets from the TMR. In this scenario, due to the projection effect there should be a spatial offset between the UV burst and the reconnection around the TMR in limb observations. Such a scenario is difficult to be confirmed, as this offset may be a few hundred km or less, which probably can not be unambiguously resolved by IRIS and current ground-based telescopes. Higher-resolution observations by future large-aperture telescopes, e.g., the Daniel K. Inouye Solar Telescope (DKIST), may tell whether this scenario is correct or not. The second proposal is that reconnection around the TMR heats the materials to different temperatures, which may be consistent with a recent numerical simulation by \cite{Ni2016}. This 2.5-dimensional reconnection model has produced various plasma components with different temperatures even in one small magnetic island. When the plasma beta is low, reconnection around the TMR can indeed heat the cool materials through small-scale shocks to a temperature of $\sim$80,000 K. \cite{Fang2017} also realized that a single heating event around the TMR can not account for the relatively long lifetime of UV bursts (minutes or even tens of minutes) due to the high radiative losses. This discrepancy may be solved if we consider continuous energy release through recurrent reconnection. Future theoretical investigations may examine whether such a process can result in relatively long-living UV bursts or not.

Figure~\ref{fig.6}~also shows a ribbon-like structure in the SJI 1400 \AA{} image. Although no X-ray flare was recorded by GOES at 18:52:12 UT, the IRIS spectra at this ribbon are similar to the typical spectra at a flare ribbon in the sense that the Si~{\sc{iv}} and C~{\sc{ii}} lines are redshifted by 15-40~km~s$^{-1}$. This suggests that a flare-like event occurred around this time, resulting in chromospheric condensation which leads to the significant red shift. The Si~{\sc{iv}} and C~{\sc{ii}} line profiles at this ribbon are much narrower and less enhanced as compared to the line profiles of the UV bursts. Unlike the complex profiles in UV bursts, the Si~{\sc{iv}} line profiles are mostly Gaussian at this ribbon. This event appears to be a microflare in the extreme-ultraviolet channels of AIA, thus largely different from the point-like UV bursts. Considering the similarity between the ribbon-like structure and typical flare ribbons, this microflare likely occurs in the chromosphere or above. It is difficult to examine whether this event is related to the UV bursts or not. However, we can imagine that frequent occurrence of small-scale reconnection in the emerging flux region may change the large-scale connectivities of magnetic field lines, thus altering the magnetic field topology of the emerging AR and affecting the triggering of large-scale solar eruptions.

\section{Evolution of photospheric magnetic fields}

\begin{figure*}
\centering {\includegraphics[width=\textwidth]{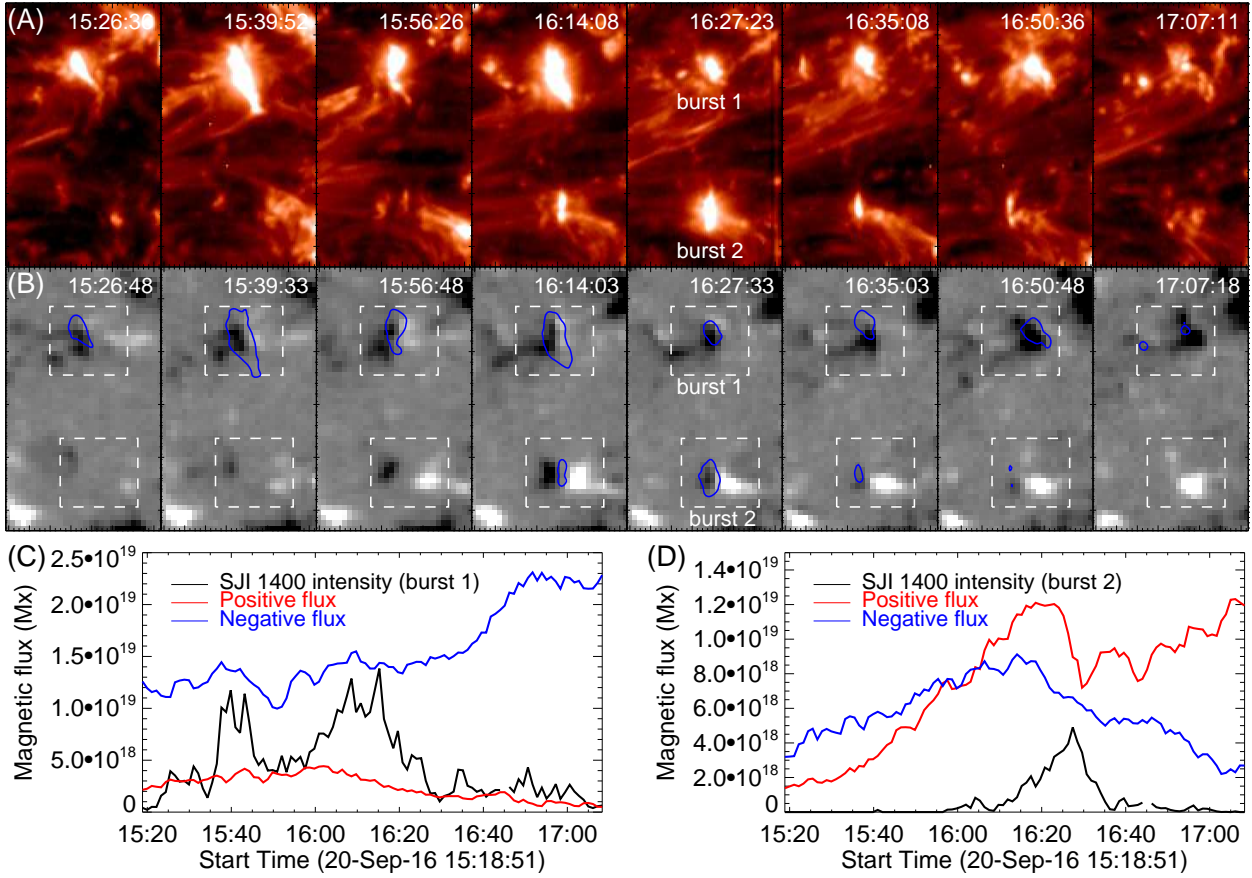}} \caption{ (A)-(B) Sequences of IRIS 1400 \AA{} images and HMI line-of-sight magnetograms in a small region enclosing bursts 1 and 2. Blue contours outlining the two UV bursts observed in the 1400 \AA{} images are overplotted in the magnetograms. The size of each image is 11 Mm$\times$18 Mm. (C)-(D) Signed magnetic fluxes integrated within the rectangular regions shown in (B) and 1400 \AA{} intensity integrated within the contours for the two UV bursts. A movie (m2.mov) is available online. } \label{fig.8}
\end{figure*}

A recent 3-dimensional radiative magnetohydrodynamic simulations of magnetic flux emergence revealed that ubiquitous reconnection between emerging bipolar magnetic fields can trigger EBs and UV bursts in the lower atmosphere \citep{Hansteen2017}. Flux cancellation is expected in such a scenario. Previous observations have shown an association of flux cancellation with a few UV bursts \citep{Peter2014,Tian2016,Nelson2016,Zhao2017}. Our observation reveals many more UV bursts and shows that flux cancellation is indeed very common for UV bursts. From the online movies we can see that most of the UV bursts are associated with evolving small-scale magnetic bipoles and that they are sitting at or around the polarity inversion lines. Many of them show clear signatures of flux cancellation, indicating the occurrence of magnetic reconnection between opposite polarities. 

Figure~\ref{fig.8}~presents the evolution of the 1400 \AA{} intensity and the line-of-sight magnetic field for two UV bursts. Burst 1 lasts for at least 2 hours and its intensity is significantly enhanced twice. During the period of 15:18 UT -- 17:08 UT, negative fluxes move towards the location of the burst from the eastern side, resulting in the roughly continuous increase of the total negative flux. At around 15:40 UT the total negative flux shows a sudden decrease, which is accompanied by an obvious enhancement of the 1400 \AA{} intensity. Some positive fluxes appear to move towards the location of the burst from the western side until 16:00 UT, leading to the increase of the total positive flux at the beginning. During the period of 16:10 UT -- 16:30 UT, flux cancellation occurs and the decrease of the total positive flux is clearly accompanied by the significant increase of the 1400 \AA{} intensity. Burst 2 reveals a more typical scenario of flux cancellation. Magnetic fluxes with opposite polarities move towards each other at a speed of $\sim$0.8~km~s$^{-1}$. Magnetic reconnection likely occurs when they meet each other at around 16:20 UT, leading to an obvious decrease in the magnetic fluxes of both polarities. In the meantime, the 1400 \AA{} intensity is greatly enhanced. If we regard the flux approaching speed as the reconnection inflow speed, and use the typical speed of the bidirectional flows $\sim$100~km~s$^{-1}$ inferred from spectroscopic observations as the reconnection outflow speed, the reconnection rate can be roughly estimated to be 0.008. This value appears to be one order of magnitude lower than the reported reconnection rates for flares \citep[e.g.,][]{Li2009,Hara2011,Tian2014b}. However, the actual inflow speed at higher layers may be significantly different from the flux approaching speed in the photosphere. In the future more studies may need to be performed to investigate whether the reconnection rates are different in the partially ionized lower atmosphere and the fully ionized corona. We have also estimated the flux cancellation rates from the slopes of the decreasing fluxes as a function of time, which turn out to be in the range of (1.5--5.0)$\times$10$^{15}$ Mx~s$^{-1}$. These values are comparable to the flux cancellation rates for UV bursts or EBs measured by \cite{Nelson2016} and \cite{Reid2016}.

\section{Magnetic field topologies}

\begin{figure*}
\centering {\includegraphics[width=\textwidth]{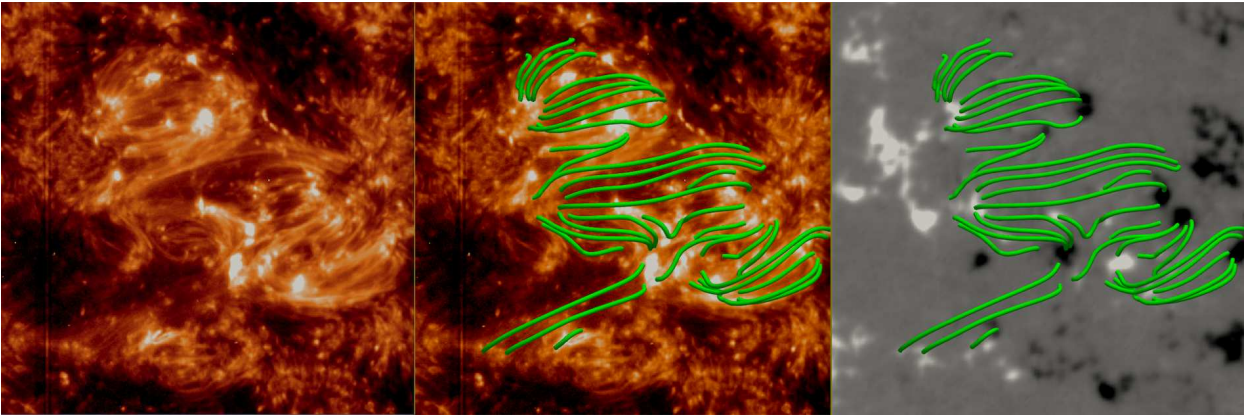}} \caption{ Left: IRIS/SJI 1400 \AA{} image taken at 16:36 UT. Middle and right: Top view of the reconstructed magnetic field lines. The background is the same 1400 \AA{} image in the middle panel and the simultaneously taken photospheric magnetogram in the right panel. } \label{fig.9}
\end{figure*}

To examine the magnetic field topologies of these UV bursts, we have reconstructed the 3-dimensional magnetic field structures from the vector magnetograms taken by HMI at 16:24 UT, 16:36 UT and 16:48 UT. Since the obtained results are similar for these three times, we only describe the analysis using the magnetogram taken at 16:36 UT. Since the UV bursts are formed in the chromosphere or even the photosphere where the force-free assumption fails, the commonly used linear or non-linear force-free magnetic field extrapolation methods may not be applicable here. Instead, we adopt a magneto-hydrostatic model through magneto-hydrodynamic(MHD) relaxation method as developed by \cite{Zhu2013} and \cite{Zhu2016}. The initial state of the relaxation comprises a sun-like plane-parallel multi-layered hydrostatic model \citep{Fan2001}, embedded with a potential magnetic field \citep{Sakurai1982}. The "stress and relax" approach \citep{Roumeliotis1996} is used, which slowly changes the transverse field at the lower boundary from potential field to the observed transverse field. Finally, the Lorentz force near the photosphere can be compensated by the pressure gradient and plasma gravity, and the magneto-hydrostatic state can be reached. In this work, the extrapolation is performed in the cubic box resolved by 608$\times$512$\times$128 grids with grid size of 0.5$^{\prime\prime}$. This grid size is the same as the pixel size of the HMI magnetograms and small enough for the investigation of magnetic topologies of UV bursts, which generally have a size of 1$^{\prime\prime}$--3$^{\prime\prime}$. Figure~\ref{fig.9}~shows a comparison between the 1400 \AA{} image taken at 16:36 UT and some magnetic field lines projected onto the plane parallel to the photosphere. Generally the cool 1400 \AA{} loops and extrapolated field lines show a good match, indicating that our extrapolation method is valid.

\begin{figure*}
\centering {\includegraphics[width=\textwidth]{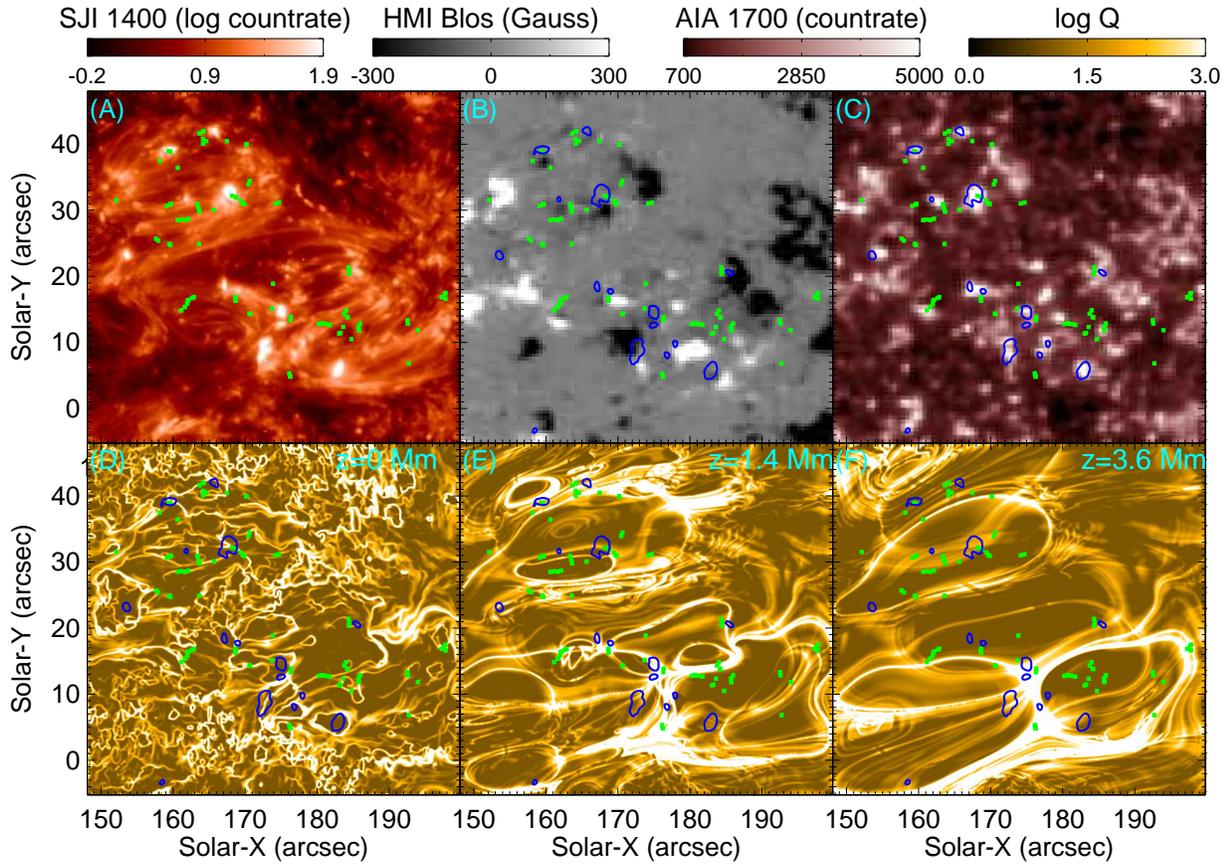}} \caption{ (A)-(C) IRIS/SJI 1400 \AA{} image, HMI line-of-sight magnetogram and SDO/AIA 1700 \AA{} image taken at 16:36 UT. (D)-(F) Images of the squashing factor at the heights of 0 Mm, 1.4 Mm and 3.6 Mm above the photosphere. Blue contours outlining the compact brightenings in the 1400 \AA{} image at 16:36 UT are overplotted in other images. The green dots indicate locations of bald patches. } \label{fig.10}
\end{figure*}

Recent investigations suggest that many UV bursts are associated with the so-called bald patches \citep{Toriumi2017, Zhao2017}. A bald patch refers to a region on the photospheirc polarity inversion line where the magnetic field line is threading through it horizontally from the negative to the positive polarity \citep{Titov1993}. It is one of the preferred locations for the occurrence of magnetic reconnection. During the flux emergence process, bald patches are often associated with the serpentine field lines formed due to the development of Parker magnetic buoyancy instability \citep[e.g.,][]{Pariat2004,Pariat2009,Xu2010,Cheung2014,Schmieder2014,Danilovic2017,Danilovic2017b}. Previous studies tend to suggest that many EBs are associated with these bald patches \citep[e.g.,][]{Georgoulis2002,Schmieder2004,Pariat2004,Centeno2017,Yang2016}. Since the curvature of the magnetic field line on the bald patch is positive, we can identify the locations of bald patches in the photosphere based on the following condition \citep{Pariat2004}:

\begin{equation}
\emph{$B_z=0~~~\rm{and}~~~\textbf{B}\cdot \nabla$$B_z$ $>$ \rm{0}}\label{equation1},
\end{equation}

where $B_z$  and $\textbf{B}$ are the longitudinal component of the magnetic field and the vector magnetic field in the photosphere, respectively. 

Since the measurement uncertainty of the HMI transverse field is $\sim$100 Gauss, we perform this calculation only at locations where the transverse field is larger than 150 Gauss. The identified bald patches are marked in Figure~\ref{fig.10}, and they are clearly located in the emerging flux region. Although a few UV bursts are located at bald patches, most UV bursts appear to be spatially offset from the bald patches. So in our observation many UV bursts may not be related to bald patch reconnection. We have also performed potential field extrapolation using only the line-of-sight component of the magnetic field data. The magnetic topology is notably different from what we have obtained using the current method. We believe that the potential field extrapolation is not suitable for the study of UV bursts occurring during processes of flux emergence. This is because in the potential field model most bald patches are found outside the emerging flux region, which is not consistent with our understanding that bald patches are very common structures in emerging flux regions \citep[e.g.,][]{Cheung2014}.

Strong electric current can form at locations where the magnetic field has strong gradients of connectivities. The squashing factor $Q$ can be used to quantify the gradients.  Locations of large $Q$ values form the so-called quasi-separatrix layers (QSLs), where magnetic reconnection often occurs. We also investigate the magnetic connectivities in the reconstructed 3-dimensional magnetic field structure by calculating the $Q$ maps at different heights above the photosphere using the method of \cite{Liu2016}. The $Q$ maps at z=0 Mm (photosphere), z=1.4 Mm and z=3.6 Mm are presented in Figure~\ref{fig.10}(D)-(F). It appears that at the height of z=1.4 Mm all UV bursts are located in regions of large $Q$ values. A similar result is also found at the height of z=0.7 Mm (not shown here). At the layers of z=0 Mm and z=3.6 Mm many UV bursts are still associated with large $Q$ values, though some bursts are found in regions of small $Q$ values. Such a result possibly suggests the most probable formation height of UV bursts around z=1 Mm, consistent with our previous suggestion that UV bursts are generated through magnetic reconnection in the lower chromosphere or upper photosphere \citep{Peter2014,Tian2016}. An analysis of the magnetic topology using the other two magnetograms taken at 16:48 UT and 16:24 UT reveals a similar result. At these two times about 50\%--70\% of the UV bursts are located at QSLs at z=3.6 Mm. However, around z=1 Mm almost all or a larger fraction of UVBs are associated with large $Q$ values. Through an investigation of the magnetic field structure, \cite{Pariat2004} concluded that EBs are produced by reconnection, not only at bald patches, but also along their separatrices. Considering the fact that many UV bursts are connected to EBs \citep{Tian2016}, our finding is consistent with this conclusion.

\section{Summary}

We have presented analysis results of intense local heating events in a unique IRIS observation of the earliest-stage flux emergence. Our results provide important constraint to the modeling of the formation and energization of active regions. 

At the beginning the region observed by IRIS appears to be a typical quiet-Sun region, showing obvious network structures in the HMI line-of-sight magnetograms~and IRIS 1400 \AA{}~images. From the photospheric 2832 \AA{}~images no sunspots can be identified in the region scanned by the IRIS slit. As time evolves, HMI observed continuous emergence of small-scale magnetic bipoles. The ongoing flux emergence is accompanied by the appearance of numerous intense transient brightenings termed UV bursts or IRIS bombs in the IRIS 1400 \AA{} and AIA 1700 \AA{} images. The total intensity of these bursts in the AIA 1700 \AA{} channel appears to stay at a relatively stable level over the course of a few hours, which may be understood as a nearly constant rate of heating by magnetic reconnection. The linear increase of the total magnetic fluxes may be the cause of the constant heating. In the meantime, discrete small patches with the same magnetic polarity move in together, causing the enhancement in the magnetic field strength. This flux merging process is accompanied by the formation of pores that is visible in the 2832 \AA{}~images. 

We have applied a single Gaussian fit to the line profiles of Si~{\sc{iv}}~1393.755 \AA{}~acquired during all raster scans. The Dopplergrams reveal significant red shifts on the order of 30~km~s$^{-1}$ at the loop footpoints and small blue shifts of a few~km~s$^{-1}$ at the loop tops, which is consistent with numerical simulations of solar flux emergence. The spectra of the UV bursts show characteristics that are similar to those of the previously reported hot explosions \citep{Peter2014}. Spectral profiles of several Si~{\sc{iv}}, C~{\sc{ii}} and Mg~{\sc{ii}} lines of these bursts are significantly broadened and enhanced relative to the normal line profiles obtained from quiet-Sun regions. Some UV bursts reveal adjacent patches of significant blue and red shifts of $\sim$100~km~s$^{-1}$, indicating the spatially resolved bidirectional reconnection outflows. Chromospheric absorption lines can also be identified from the greatly enhanced wings of the Si~{\sc{iv}} and C~{\sc{ii}} lines,  suggesting heating of the cool plasma to a few tens of thousands of kelvin by magnetic reconnection in the lower chromosphere or even the photosphere.  

Most of the UV bursts appear to result from interactions between magnetic fields with different polarities. Flux cancellation can be clearly identified for many bursts. By inputing the vector photospheric magnetograms to a magneto-hydrostatic model, we have reconstructed the three-dimensional magnetic field structures for the UV bursts. Different from previous investigations \citep{Toriumi2017, Zhao2017}, we find that only a small fraction of these bursts are associated with the bald patch configuration. We have also investigated the magnetic connectivities for the UV bursts, and find that almost all bursts are located in regions of large squashing factors around the height of z=1 Mm. Since large squashing factors are a reflection of strong gradients of magnetic connectivities and strong electric current, this result supports the suggestion that UV bursts are powered by small-scale magnetic reconnection in the lower solar atmosphere. Considering their coincidence with not only bald patches but also separatrices, these reconnection events are similar to Ellerman bombs \citep{Pariat2004}.

\begin{acknowledgements}
IRIS is a NASA Small Explorer mission developed and operated by LMSAL with mission operations executed at NASA Ames Research center and major contributions to downlink communications funded by ESA and the Norwegian Space Center. This work is supported by NSFC grants 41574166, 11790304 (11790300), 11403044 and 11503089, the Recruitment Program of Global Experts of China, the Max Planck Partner Group program, and the Strategic Pioneer Program on Space Science, Chinese Academy of Sciences, Grant No. XDA15011000 and XDA15010900. H. Tian and H. Peter acknowledge the support by ISSI to the team "Solar UV bursts - a new insight to magnetic reconnection". We thank Dr. Rui Liu for helpful discussion and the anonymous reviewer for constructive suggestions.
\end{acknowledgements}

\end{document}